\documentclass[aps,twocolumn,showpacs,amsmath,amssymb]{revtex4}
\usepackage[T2A]{fontenc}
\usepackage{amsmath}
\usepackage{amssymb}
\usepackage{graphicx}
\usepackage{euscript}

\begin{document}
\title{Detecting barrier to cross-jet Lagrangian transport 
and its destruction in a meandering flow}

\author{M.V. Budyansky, M.Yu. Uleysky, and S.V. Prants}
\affiliation{Pacific Oceanological Institute \\
of the Russian Academy of Sciences, 43 Baltiiskaya st., 690041 Vladivostok,
Russia}
\begin{abstract}
Cross-jet transport of passive scalars in a kinematic model of the 
meandering laminar two-dimensional incompressible flow  which is known to 
produce chaotic mixing is studied. We develop a method for detecting 
barriers to cross-jet transport in the phase space which is a physical space for 
our model. Using tools from theory of nontwist maps, we construct 
a central invariant curve and compute its characteristics that may serve 
good indicators of the existence of a central transport barrier, its strength, 
and topology. Computing fractal dimension, length, and winding number of that curve 
in the parameter space, we study in detail change of its geometry and its 
destruction that are caused by local bifurcations and a global bifurcation 
known as reconnection of separatrices of resonances. Scenarios of reconnection 
are different for odd and even resonances. The central invariant curves 
with rational and irrational (noble) values of winding numbers are 
arranged into 
hierarchical series which are described in terms of continued fractions. 
Destruction of  central transport barrier is illustrated for two ways 
in the parameter space: when moving along resonant bifurcation curves with 
rational values of the winding number and along curves with   
noble (irrational) values. 
\end{abstract}
\maketitle

\section{Introduction}
A meandering jet is a fundamental structure in laboratory
and geophysical fluid flows. Strong oceanic and atmospheric jet
currents separate water and air masses with distinct physical properties.
For example, the Gulf Stream separates the colder and fresher slope
ocean waters from the salty and warmer Sargasso sea ones.
Recently, there has been
much interest in applying ideas and methods from dynamical systems
theory to study mixing and transport in meandering jets. In
steady horizontal velocity fields, water (air) parcels move along
streamlines in a regular way. When the velocity field changes in time,
the motion becomes much more complicated even if the change is periodic.
The phenomenon of {\it chaotic advection} of passive particles
in (quasi)periodically-disturbed fluid flows has been studied theoretically
and experimentally \cite{A65,A84,A02,Ottino}.

In the context of dynamical systems theory, chaotic advection is
Hamiltonian chaos in two-dimensional incompressible flows (for a
review of Hamiltonian chaos see, for example,
\cite{AKN85,LL84,Ott,Z05}). The coordinates $x$ and $y$ of a passive 
particle on the horizontal plane satisfy to simple Lagrangian
equations of motion 
\begin{equation}
\frac{d x}{d t}=u(x,y,t)=-\frac{\partial\Psi}{\partial y},\quad
\frac{d y}{d t}=v(x,y,t)=\frac{\partial\Psi}{\partial x},
\label{eqn1}
\end{equation}
where $u$ and $v$ are zonal and meridional velocities of the particle, and
the stream function $\Psi$ plays the role of a Hamiltonian. The phase
space of the dynamical system (\ref{eqn1}) with one and half degrees of 
freedom is a configuration space for passive particles.

A number of simple kinematic and dynamically consistent model stream functions
have been proposed to study large-scale chaotic mixing and transport
in geophysical meandering jet flows \cite{S92,DW96,Miller97,YPJ02,Chaos06,
UlBP07,PK06,Book08}. Deterministic models do not pretend to quantify
transport fluxes in real oceanic and atmospheric currents but they are
useful to reveal large-scale space-time structures that specify
qualitatively mixing and transport of water and air masses. Whether or not the jet
provides an effective barrier to meridional or
{\it cross-jet transport},
under which conditions the barrier becomes permeable and
to which extent, these are crucial questions in physical
oceanography and physics of the atmosphere.
The problem must be treated from different points of view. In the
straightforward numerical approach based on full-physics nonlinear models,
the velocity field is generated as an outcome of a basin circulation
model and a flux across the jet (if any) can be estimated integrating
a large number of tracers. The kinematic and linear dynamically
consistent models are less realistic, but they allows to identify
and analyze different factors which could enhance or suppress the
cross-barrier transport. 

As to meandering currents, both the approaches have been 
applied to study cross-jet Lagrangian transport. A simple
kinematic model with the basic streamfunction in the form
(\ref{eqn2}) has been shown to reproduce some features of the
large-scale Lagrangian dynamics of the Gulf Stream water masses
\cite{B89}. The phase portrait of Eqs.~(\ref{eqn1}) with the
meandering Bickley jet (\ref{eqn2}) is plotted in Fig.~\ref{fig1}~(a)
in the frame moving with the meander phase velocity.
Time dependence of the meander amplitude or the introduction of a
secondary meander, superimposed on the basic flow, may break
the boundaries between distinct regions in Fig.~\ref{fig1}~(a)
producing chaotic mixing and transport between them
\cite{S92, Meyers94,DW96,Chaos06,UlBP07,JPA08}. The numerical
calculations, based on computing the Melnikov function \cite{Melnik},
have shown that transport across the jet was much weaker than that
between the jet ($J$), the circulation cells ($C$), and
the peripheral currents ($P$) in Fig.~\ref{fig1}~(a), i.~e.
the perturbation mixes the water along each side of the jet more
efficiently than across the jet core \cite{S92}. An attempt
to analytically predict the parameter values for the destruction
of the transport barrier was made in Ref.~\cite{Meyers94} using
the heuristic Chirikov criterion for overlapping resonances
\cite{Chir79}. A technique, based on computing the finite-scale
Lyapunov exponent, as a function of initial position of tracers,
has been found useful in Ref.~\cite{BLR01} to detect the presence of
cross-jet barriers in the kinematic model (\ref{eqn2}).
An analysis of cross-jet transport, based on lobe dynamics, has been 
applied in \cite{RWig06} to describe how particles can cross the jet
from the north to the south and vice versa.

The study of cross-jet transport has been motivated also by a series
of laboratory experiments \cite{SMS89,BMS91,SHS93} on Rossby waves
propagating along an azimuthal jet in a rapidly rotating tank.
This flow can be modeled in the linear approximation of the corresponding
fluid equations \cite{DM93} by a stream function which is a superposition
of a Bickley jet and two neutral modes (Rossby waves). The destruction
of a barrier to cross-jet transport has been studied analytically by
using the Chirikov criterion in the pendulum approximation and
numerically by using Poincar{\'e} sections \cite{DM93}.
It was shown that one needs very large values of the perturbation amplitudes
to break the barrier. The analytic model proposed in \cite{DM93}
has been used recently to study Lagrangian dynamics of atmospheric zonal jets
and the permeability of the stratospheric polar vortex. Poincar{\'e}
sections and finite-time Lyapunov exponents revealed a robust transport
barrier which can be broken either due to large perturbation 
amplitudes of the Rossby waves or as a result of an increase 
of their phase velocities \cite{Rypina}. A comparison of properties
of cross-jet transport in ad hoc kinematic and dynamically consistent
models of atmospheric zonal jets has been done recently in
Ref.~\cite{P.H.Haynes}.

Being motivated by Lagrangian observations of the oceanic currents, cross-jet
transport and mixing have been studied in numerical models of meandering
jets \cite{Miller97,YPJ02,YPJ04}. It has been shown both in barotropic
and baroclinic nonlinear numerical models, where the meander amplitude can not 
be made arbitrary large, that cross-jet chaotic transport, resulting from the
meandering motions, are maximized at a subsurface level. Since the
undisturbed velocity is weaker at deeper levels, the corresponding
separatrices are closer to the jet core. Therefore, separatrix reconnection
should occur below some critical depth and transport across the jet should
be facilitated. 

Independent on the work on cross-jet transport in the geophysical community,
there have been a number of theoretical and numerical investigations
of chaotic transport in so-called area-preserving nontwist maps
\cite{HH84,W88,DM93,PhysD96,Aizawa,PhysD97,Shinohara97,Wurm05,Wurm04}.
We mention specially the early study of different
reconnection scenario \cite{HH84, W88} and the first
systematic study of cross-jet transport in nontwist
maps \cite{DM93,PhysD96,PhysD97}.
These maps locally violate the twist condition, a map analogue of the 
non-degeneracy condition for Hamiltonian systems. Nontwist maps are of 
interest because many important mathematical results, including  
KAM and Aubry-Mather theory, depend on the twist condition. Apart from
their mathematical importance, nontwist maps are of a physical interest
because they are able to model transition to global chaos, the term
meaning in the mathematical community a cross-jet transport. Nontwist
maps allow to study different scenarios for this transition:
reconnection of separatrices, meandering and breakup of invariant
tori, and others.

The onset of global chaos in oscillatory Hamiltonian systems, where the
eigenfrequency possesses a local extremum as a function of energy,
has been studied analytically and numerically in
Refs.~\cite{SMSoskin,S.M.S}.
In such systems with two or more separatrices, global chaos may occur 
at unusually small magnitudes of perturbation due to overlap
in the phase space between resonances of the same order and
their overlap in energy with chaotic layers of the corresponding
unperturbed separatrices.

In the present paper we develop a method for detecting a barrier to cross-jet
Lagrangian transport (or global chaos in a more general context),
apply it to the kinematic model of a meandering jet flow,
study changes in its topology under varying the perturbation
parameters, and scenarios of its destruction. In Sec.~II we briefly
introduce a model streamfunction which is known to produce
chaotic advection \cite{S92,Chaos06,RWig06} and
compute the amplitude--frequency $\varepsilon$--$\nu$ diagram 
demonstrating the parameter range for
which cross-jet transport exists. Based on the symmetry
of the flow, we propose in Sec.~IIIa a numerical method to identify
{\it a central invariant curve} (CIC) which is a diagnostic means to detect 
the process of destruction of {\it a central transport barrier} (CTB). 
The CIC is constructed by successive iterations of so-called indicator points 
\cite{Aizawa}. Computing the fractal dimension of a set of 
iterations of those points at different values of the parameters, 
we identify whether the CIC and CTB are broken or not. 
In Sec.~IIIb we study possible geometries of the CIC
that may change dramatically with varying $\varepsilon$
and $\nu$. Before the total destruction, the CIC
experiences a number of local bifurcations becoming a complicated
meandering curve whose properties can be specified by its length
and the winding number $w$.
A structure of the set of CICs is revealed in  a continued fraction representation 
of their winding numbers. The CICs with rational $w$ are 
arranged in hierarchical series connected with the corresponding resonances. 
Whereas, the CICs with noble numbers form their own series.  
Destruction of CTB is studied in Sec.~IV for two ways in the parameter space. 
When moving along a so-called resonant bifurcation curve with a rational value 
of $w$, one specifies the values of $\varepsilon$ and $\nu$ for which the CIC 
is broken but CTB remains. In contrary to that, when moving along any curve 
with noble value of $w$, a CIC exists providing CTB. The process of 
CTB destruction in both the cases is illustrated in Sec.~IV.

\section{The amplitude--frequency diagram for cross-jet transport in the
model flow}

\begin{figure}[!htb] 
\includegraphics[width=3.4in,clip]{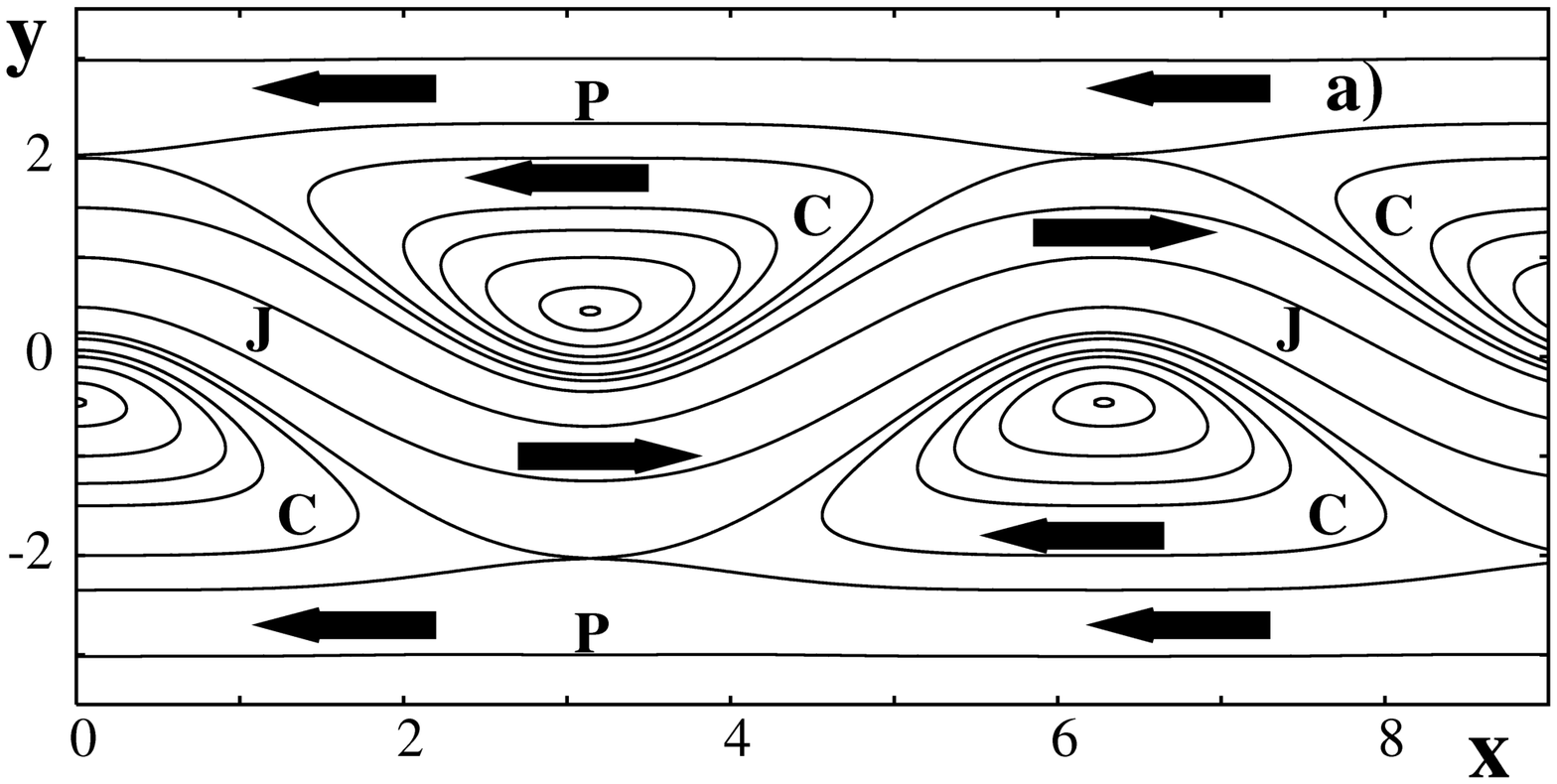}\\
\includegraphics[width=3.4in,clip]{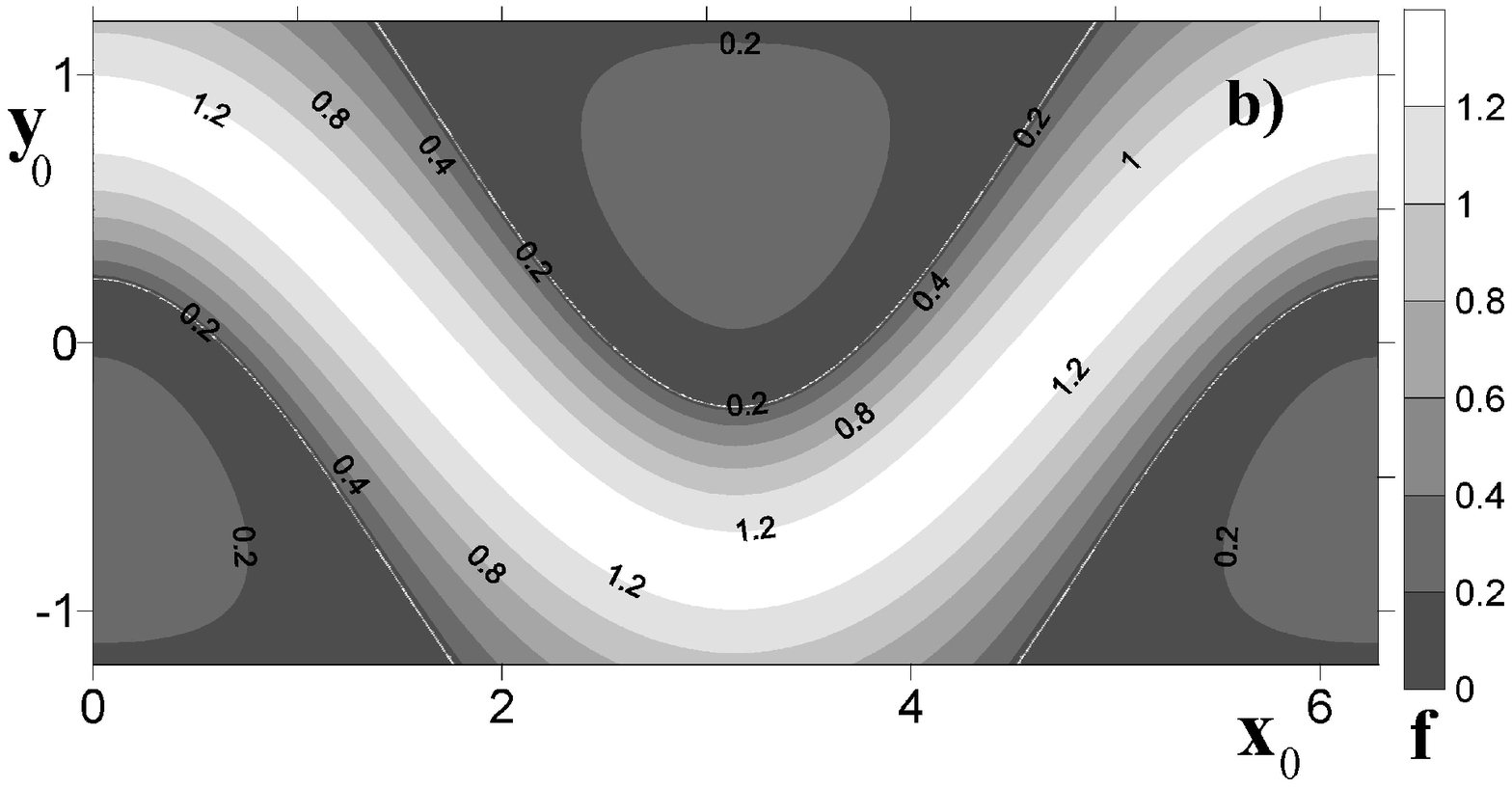}
\caption{(a) Phase portrait of the unperturbed flow
in the frame moving with the meander's phase velocity.
Streamlines in the circulation cells ($C$),
jet ($J$), and peripheral currents ($P$) are shown.
(b) Frequency map represents by color values
of the frequency $f$ of particles with initial positions
($x_0, y_0$) advected by the unperturbed flow.} 
\label{fig1} 
\end{figure}

We take the Bickley jet with a running wave imposed as a kinematic model
of a meandering shear flow in the ocean. The respective 
normalized stream function in the frame moving with the phase
velocity of the meander has the following form \cite{Chaos06}:
\begin{equation}
\Psi=-\tanh{\left(\frac{y-A\cos x}{L_{\text{jet}}\sqrt{1+A^2\sin^2 x}}\right)}+Cy,
\label{eqn2}
\end{equation}
where the jet's width $L_{\text{jet}}$, meander's amplitude $A$ and its phase
velocity $C$ are the control parameters. The phase portrait of the advection
equations (\ref{eqn1}) with the streamfunction (\ref{eqn2}),
shown in Fig.~\ref{fig1}~(a), consists of three different regions:
the central eastward jet $J$, chains of the northern and southern
circulation cells $C$ and the peripheral westward currents $P$.
The flow is steady in the moving frame of reference, and passive
particles follow the streamlines. 
In Fig.~\ref{fig1}~(b) we plot a frequency map $f(x_0,y_0)$ that shows
by nuances of the grey color the value of the frequency $f$ of particles with initial
positions ($x_0,y_0$) in the unperturbed system. The maximal value of    
the frequency, $f_{\text{max}}=1.278$, have the particles moving in the
central jet. 

As a perturbation, we take the simple periodic modulation of the
meander's amplitude
\begin{equation}
A=A_0+\varepsilon\cos \nu t.
\label{eqn3}
\end{equation}
Under the perturbation, the separatrices, connecting saddle
points, are destroyed and transformed into stochastic layers.
The strength of chaos depends strongly on both the perturbation
parameters, the perturbation amplitude $\varepsilon$ and frequency $\nu$.
In the model used the normalized control parameters are connected with the
dimensional ones as follows \cite{Chaos06}: $A=ak$, $C=c/u_{\text{m}}\lambda k$,
and $L_{\text{jet}}=\lambda k$, where $a,k$, and $c$ are amplitude, wave
number, and phase velocity of a meander, respectively,
$\lambda$ and $u_{\text{m}}$ are characteristic width and maximal zonal velocity
in the jet on the surface. All these parameters change in a wide range
in the Gulf Stream \cite{B89,S92,Meyers94}: $\lambda\simeq 40\div 100$~km,
$a\simeq 50\div 60$~km, $2\pi/k\simeq 200\div 400$~km,
$c\simeq 0.1\div 0.5$~m/sec, $u_{\text{m}}\simeq 1\div1.5$~m/sec.
So, we get $L_{\text{jet}}\simeq 0.1\div 3$, $A\simeq 0.7\div 2$, and
$C\simeq 0.02\div 0.3$.
Being motivated by mixing and transport in the Gulf Stream,
we took the following normalized values of the control parameters that
will be used in
all our numerical experiments: $A_0=0.785, C=0.1168$ and $L_{\text{jet}}=0.628$.

The equations of motion (\ref{eqn1}) with the stream function 
(\ref{eqn2}) and the perturbation (\ref{eqn3})
have the symmetry 
\begin{equation}
\hat S:\left\{
\begin{aligned}
x'&=\pi+x,\\
y'&=-y
\end{aligned}\right.
\label{*}
\end{equation}
and the time reversal symmetry
\begin{equation}
\hat I_0:\left\{
\begin{aligned}
x'&=-x,\\
y'&=y.
\end{aligned}\right.
\label{**}
\end{equation}
The symmetries (\ref{*}) and (\ref{**}) are involutions, i.~e.
$\hat S^2=1$ and $\hat I_0^2=1$.
Due to the symmetry $\hat S$, motion can be
considered on the cylinder with $0\le x\le 2\pi$.
The part of the phase space with $2\pi n\le x\le 2\pi (n+1)$,
$n=0, \pm 1, \dots$, is called a frame. It should be stressed that the
phase space in two-dimensional incompressible flows is a configuration
space for advected particles.

\begin{figure}[!htb] 
\includegraphics[width=3.4in,clip]{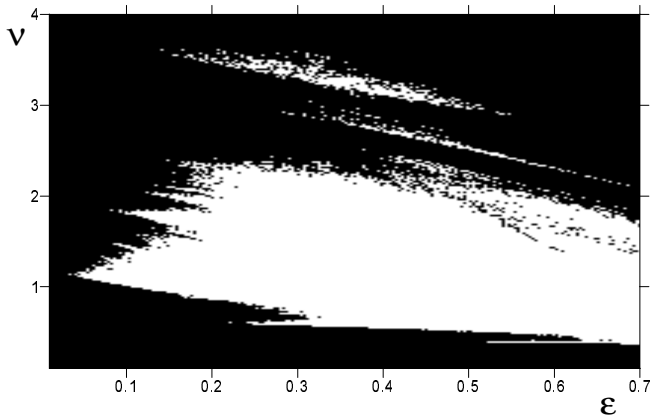}
\caption{Amplitude--frequency $\varepsilon$--$\nu$ diagram  
showing the parameter values for which 
cross-jet chaotic transport exists (white zones) or not
(black zones).} 
\label{fig2} 
\end{figure}

The following numerical procedure has been applied to establish the fact of 
cross-jet transport in the kinematic model of the meandering
jet current. The advection equations (\ref{eqn1}) for given values of the
perturbation
amplitude $\varepsilon$ and frequency $\nu$ and with twenty particles,
released nearby the northern saddle point, have been integrated up to the
time instant when one of the particles was detected to cross
the straight line $y=y_s$ passing through the southern saddle.
If after the time $T_{\text{max}}=1000\times 2\pi/\nu$
none of the particles crosses the line $y=y_s$, we assume that for
given values of the parameters there is no cross-jet transport.
The $\varepsilon$--$\nu$ diagram in Fig.~\ref{fig2} shows the
values of the parameters for which cross-jet chaotic transport exists
(white-color zones). There are a number of the frequency values for
which transport occurs at surprisingly small values of the perturbation
amplitude $\varepsilon$. 
The absolute minimal value of the perturbation amplitude, at which  
the cross-jet chaotic transport occurs, $\varepsilon_{\text{min}}=
0.0218\approx A_0/36$, corresponds to the frequency $\nu=1.165$ which
is close to the natural frequencies of the particles moving in the central jet. 

\section{Topology of a barrier to cross-jet transport and central invariant curve}

The amplitude--frequency diagram is useful to detect cross-jet transport but
its computation is a time consuming procedure. Moreover, it says
nothing about the properties of barrier to transport and mechanism
of its destruction. A fractal-like boundary between the colors in 
Fig.~\ref{fig2} reflects an intermittency
in appearance and destruction of the cross-jet barrier when varying 
$\varepsilon$ and $\nu$. Further insight into topology
of the barrier could be obtained if one would be able to
find an indicator 
of cross-jet transport, i.~e. an object in the phase space whose 
form contains an information about permeability of the barrier.

\subsection{Detecting the central invariant curve}

First of all, we need to give definitions of some basic
structures specifying a cross-jet barrier and its 
destruction.
{\it The central transport barrier} (CTB) is defined
as a strip between the southern and northern
unperturbed separatrices confined by marginal
northern and southern ballistic trajectories
(excluding orbits of ballistic resonances in the stochastic layer).
All the trajectories inside the CTB are ballistic, some of them are
regular and the other ones are chaotic.
The amplitude--frequency diagram in Fig.~\ref{fig2}
demonstrates clearly destruction of the CTB at some values of
the perturbation parameters $\varepsilon$ and $\nu$ and
onset of cross-jet transport.

Our Hamiltonian flow with the streamfunction (\ref{eqn2}) is degenerate,
i.~e. it violates the non-degeneracy condition, 
$\partial f/\partial I\ne 0$, for some values of the natural
frequency of passive particles $f$ 
and their actions $I$ in the unperturbed system. 
Physically it means that the zonal velocity profile $u(y)$
has a maximum. In theory
of nontwist maps the curve, for which the twist condition (analogue of 
the non-degeneracy condition)   
is violated, is called a nonmonotonic curve \cite{Wurm05}.
In our model flow (\ref{eqn2}) it is some value of unperturbed
streamfunction along which the frequency $f$
is maximal (see Fig.~\ref{fig1}~(b)).

Instead of integrating the advection equations (\ref{eqn1}), we integrate
the corresponding Poincar{\'e} map, an orbit of which is defined as
a set of points $\{(x_i,y_i)\}_{i=-\infty}^{\infty}$ on the phase
plane such that
$\hat{G}_{T}(x_i, y_i)=(x_{i+1},y_{i+1})$, where $\hat{G}_t$ is an
evolution operator on a time interval $t$ and $T\equiv 2\pi/\nu$
is the period of perturbation. Operator $\hat{G}_{T}$ can be factorized
as a product of two involutions $\hat{G}_{T}=\hat I_1\hat I_0$, where
$\hat I_1=\hat G_T\hat I_0$~ is also a time reversal
simmetry~\cite{PhysD96}.

A periodic orbit of period $nT$ ($n=1,2,\dots$) is an orbit such that
$(x_{i+n}, y_{i+n})=(x_i+2\pi m,y_i)$, $\forall i$, where $m$ is an integer.
An invariant curve is a curve invariant under the map.
The nonmonotonic curve is not an invariant curve under a perturbation.
The winding (or rotation) number $w$ of an orbit is defined as the limit
$w=\lim\limits_{i\to \infty}[(x_i-x_0)/(2\pi i)]$, when it exists.
The winding number is a ratio between the frequency of
perturbation $\nu$ and the natural frequency $f$.
Periodic orbits have rational winding numbers $w=m/n$. It simply means 
that a ballistic passive particle in the flow flies $m$  frames before 
returning to its initial position $x_0$ (modulo $2\pi$) after $n$ periods of 
perturbation. Winding numbers of quasiperiodic orbits are irrational. 

Now we are ready to introduce the important notion of 
{\it a central invariant curve} (CIC). We define CIC as
an curve which is invariant under the operators $\hat S$ and $\hat G_T$.
It can be shown, that two curves invariant under $\hat S$ have at
least two common points. The curves, which are invariant under $\hat G_T$, 
cannot intersect each other. So, the CIC is a unique curve.
Following to Ref. \cite{Shinohara97}, 
one can show that the CIC corresponds to a local extremum on the 
winding number profile with an irrational value of $w$.
Such curves are called <<shearless curves>> in theory of nontwist maps \cite{Wurm05}.
The significance of a shearless curve is that it acts as a
barrier to global transport in the phase space of a nontwist map.
The violation of the twist condition leads to existence of more
than one orbit with the same winding number  
arising in pairs on both sides of the shearless curve.
Those pairs of orbits can collide and annihilate at certain
parameter values. The collision of the orbits involves in
phenomenon, which was called as reconnection of 
invariant manifolds of the corresponding hyperbolic orbits \cite{HH84}. 

The CIC should not be thought as the last cross-jet barrier curve in the
CTB in the sense that it breaks down under increasing the perturbation
amplitude in the last turn. Sometimes it is the case, but sometimes it is not. 
Nevertheless, the CIC serves a good indicator of the strength of the CTB
and its topology.

The CIC can be constructed by successive iterations of so-called
indicator points \cite{Aizawa}. In our model flow (\ref{eqn2})
with the symmetries (\ref{*}) and (\ref{**}), indicator points are
the points ($x_j^{(k)}$, $y_j^{(k)}$), $k=1,2$, which are solutions of
the equations
\begin{equation}
\hat I_0(x_j^{(1)}, y_j^{(1)})=\hat S(x_j^{(1)}, y_j^{(1)}),
\label{ind1}
\end{equation}
or
\begin{equation}
\hat I_1(x_j^{(2)}, y_j^{(2)})=\hat S(x_j^{(2)}, y_j^{(2)}),
\label{ind2}
\end{equation}
where index $j$ numerates the points.
The equation (\ref{ind1}) gives a pair of indicator points:
($x_1^{(1)}=\pi/2$, $y_1^{(1)}=0$) and ($x_2^{(1)}=3\pi/2$, $y_2^{(1)}=0$).
Instead of solving Eq. (\ref{ind2}) we solve the equivalent equation
\begin{equation}
\hat G_T(x,y)=\hat I_0\hat S(x,y)\equiv(\pi-x,-y).
\label{ind3}
\end{equation}
If some ($x$, $y$) is a solution of (\ref{ind3}), then $\hat I_0(x$, $y)$~ 
is a solution of (\ref{ind2}). The equation (\ref{ind3}) cannot be solved
analitically, so we apply the numerical method based on computing
a minimum of the function $r(x,y)=||\hat{G}_T(x,y)-(\pi-x,-y)||$,
where $||\cdot||$ is a norm on the cylinder.
Since $r(x,y)\ge 0$ for any $(x,y)$, the points with $r(x,y)=0$ 
are minima of the function $r(x,y)$. Thus, solution of Eq. (\ref{ind3})
reduces to searching for a local minimum of the function
$r(x,y)$ with the additional condition $r(x,y)=0$ at the
point of the minimum. There are a number of numerical methods for
doing that job. We prefer to use the downhill simplex method.
In our problem the function $r(x,y)$ always has two minima, i.~e.
two indicator points transforming to each other under
action of the operator $\hat S$.

Next, we study iterations, i.~e. Poincar{\'e} mapping of one of
the indicator points $(x_0,y_0)$.
If the iterations $(x_i,y_i)=\hat G_T^i(x_0,y_0)$ are confined between
invariant curves in a
bounded region, the following three cases are possible in dependence
on the dimension $d$ of the set $(x_i,y_i)$:

1) The iterations lie on a curve on the phase plane with $d=1$ 
which is a CIC. 

2) The iterations is an organized set of points with $d=0$.
It means that they constitute either a central periodic orbit or 
a central almost periodic orbit, an orbit that could not form 
a smooth curve on the phase plane for a limited integration time.  

3) The iterations form a central stochastic layer with $d=2$.

If the iterations are not confined by any invariant curves in a
bounded region, i.~e. they occupy all the accessible phase plane
to the south and north from the central jet, then there exists 
global chaotic transport. 
Thus, the type of motion of indicator points provides an
indicator of global chaos and the absence of barriers to cross-jet transport.

\begin{figure}[!htbp]
\centerline{\includegraphics[width=3.4in]{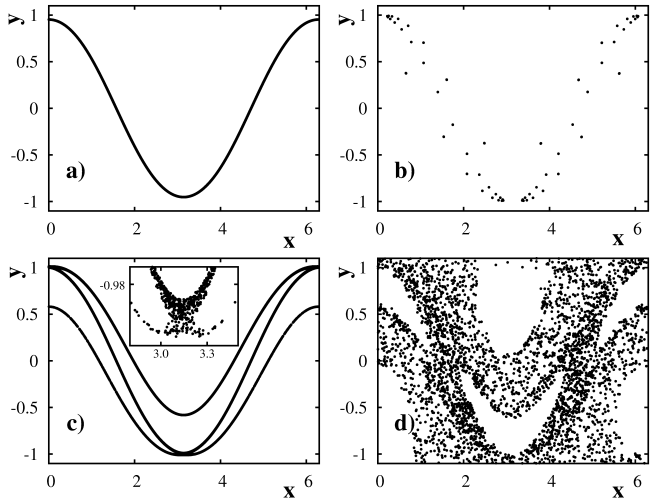}}
\caption{Poincar{\'e} mapping of indicator points.
In the first three panels the orbit of these points
is bounded and there exists a central transport 
barrier (CTB) with (a) CIC $(\varepsilon=0.01,\nu=1.2)$,
(b) central almost periodic orbit $(\varepsilon=0.011997277,\nu=1.2)$,
and (c) central stochastic layer at $(\varepsilon=0.01177,\nu=1.2)$ 
with the inset demonstrating a magnification of a small region. 
(d) Destruction of CTB and onset of global chaotic transport as a result of 
unbounded iterations of indicator points $(\varepsilon=0.041,\nu=1.2)$.}
\label{fig3}
\end{figure}

Possible topologies of the corresponding CTB at the fixed frequency $\nu=1.2$ 
and with increasing values of 
the perturbation amplitude $\varepsilon$ 
are illustrated in Fig.~\ref{fig3} plotting iterations of the 
indicator points computed by the above-mentioned method. 
The panel (a) illustrates the case when those iterations form a CIC. 
Another typical situation is shown in Fig.~\ref{fig3}~(b) where the iterations 
fall in small segments  filling at $\tau \to \infty$ a continuous curve
which is a central almost periodic orbit. If the iterations fill up not a curve
but a bounded region 
between invariant curves, then there appears a central stochastic layer 
preventing cross-jet transport (Fig.~\ref{fig3}~(c)). 
When iterations of the indicator points occupy 
a region that is not confined by any  invariant curves, it means
destruction of the CTB and onset of global chaos, i.~e., chaos in a large region 
of the phase space accompanied by cross-jet transport. 

The indicator points have been found with the help of the above-mentioned 
numerical procedure,  
and their iterations have been computed in the following range of the control
parameters: $\nu\in[0.95:1.5]$ and $\varepsilon\in [0.01:1]$.
We assume that the iterations are bounded, if their
coordinates do not cross the unperturbed separatrices after
$5\times 10^4$ iterations. The dimension $d$ of the set of those 
iterations is computed by the
box-counting method, where the value of $d$ for the box size $e_k=(1/2)^k$
is defined as
\begin{equation}
d_k=\log_2\frac{N_{k+1}}{N_k},
\label{eqn4}
\end{equation}
where $N_k$ is a number of boxes of the size $e_k$ containing set points.
The dimension $d_k$ goes to zero with decreasing $e_k$,
and one cannot distinguish in this limit between the central
almost periodic
orbit and the central stochastic layer at large $k$.
Comparing the values of $d_k$ at different values of $k$, 
we were able to find the empirical value $k=4$
which is enough to make the difference.

\begin{figure}[!htbp]
\centerline{\includegraphics[width=3.4in]{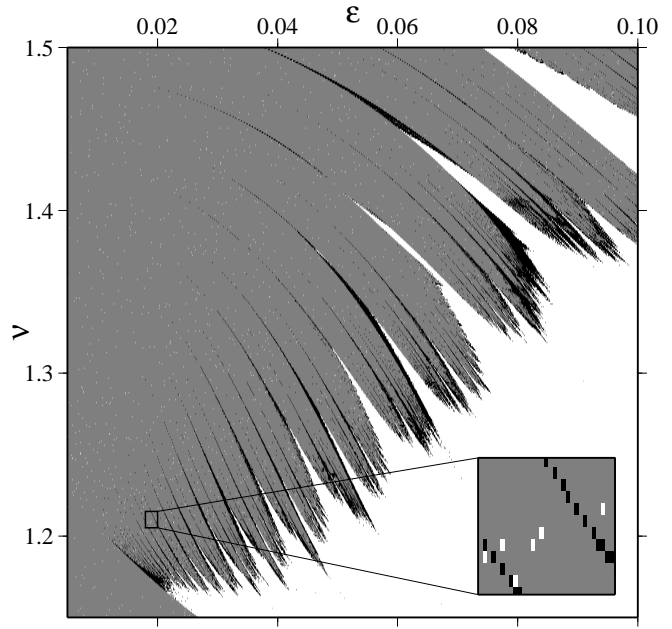}}
\caption{Bird-wing diagram of the box-counting dimension $d_4(\varepsilon,\nu)$. 
White color: regime with global chaotic transport with 
unbounded motion of indicator points (see Fig.~\ref{fig3}~(d)).
If the motion of indicator points 
is bounded, then  there exists a CTB but its topology may differ. 
Grey color ($0.95 \le d_4 \le 1.05$):
regime with a CIC (see Fig.~\ref{fig3}~(a)).   
Small white regions which are hardly visible inside the grey ``wing''
($d_4<0.95$): 
regime with a central almost periodic orbit (see Fig.~\ref{fig3}~(b)). 
Black color ($d_4>1.05$): regime with a central stochastic layer 
(see Fig.~\ref{fig3}~(c)). 
Inset shows magnification of a small region in the parameter space 
with visible white and black regions.} 
\label{fig4}
\end{figure}

The results of computation of the dimension  
$d_4(\varepsilon,\nu)$ for a set of iterations of the indicator points 
are shown in the bird-wing diagram in Fig.~\ref{fig4}. 
That one and the other bird-wing diagrams in the parameter space 
show the properties of CTB and CIC in the range of comparatively small values 
of the perturbation amplitude ($0.01 \le \varepsilon \le 0.1$) and the frequency 
($1.15 \le \nu \le 1.5$) corresponding to particles moving in the 
central jet (Fig.~\ref{fig1}~(b)). 
White color corresponds to the regime of global chaotic transport with 
unbounded motion of iterations of the indicator 
points. Otherwise, the CTB exists but its topology is 
different. Grey color means that there exists a CIC with 
$0.95 \le d_4 \le 1.05$  
in the corresponding range of the parameters.  
White rectangles, which are hardly visible in the main panel 
(see their magnification on the inset of the figure), means existence of 
a central almost periodic orbit with $d_4<0.95$ and 
black strips~--- a central stochastic layer with $d_4> 1.05$.

\subsection{Geometry of the central invariant curve and its bifurcations}

To quantify complexity of the CIC 
we define its length $L$ as a sum of the distances between the 
iterations of the indicator points $(x_i,y_i)$ ordered 
on the phase plane in the following way:
\begin{enumerate}
\item The first step. A point $B_0$, belonging to a set of iterations of the
Poincar{\'e} map $(x_i,y_i)$, is marked.
\item The $(j+1)$-th step. We find and  mark among all the unmarked points 
that one, $B_{j+1}$, which minimizes the Euclidean distance 
$D_j=D(B_j,B_{j+1})$ between $B_j$ and $B_{j+1}$.
\item The procedure is repeated unless all the points will be marked.
\end{enumerate}

As an output we have an ordered set of points $B_j$ constituting a CIC.
The accuracy is controlled by the quantity $\max{D_j}$.
Large values of this quantity mean that the points
are ordered in a wrong way or a set of points is chaotic.
To increase the number of points we use in addition to the
original points $(x_i,y_i)$ their ``images''
$(x_i+\pi,-y_i)$ as well. To minimize the computation time
the points are sorted in accordance with their $x$ coordinates.

\begin{figure}[!htbp]
\centerline{\includegraphics[width=3.4in]{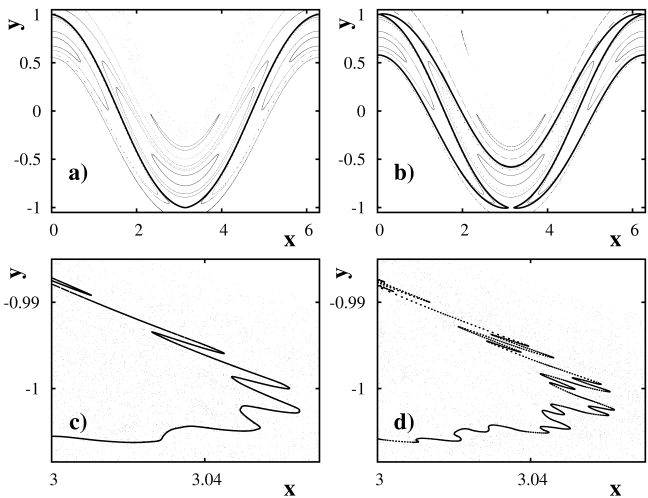}}
\caption{Metamorphosis of the central invariant curve (CIC). 
(a) Nonmeandering CIC ($\nu=1.2$, $\varepsilon=0.01174929$).
(b) Meandering CIC of the first order and period $T$
($\nu=1.2$, $\varepsilon=0.01178721$). 
(c) Meandering CIC of the second order and period $79T$ 
($\nu=1.2$, $\varepsilon=0.01179027$).
(d) Meandering CIC of a higher order ($\nu=1.2$,
$\varepsilon=0.01179339$).}
\label{fig5}
\end{figure}

Figure \ref{fig5} illustrates metamorphosis of the CIC 
as the perturbation amplitude increases.
We start with the CIC, shown in Fig.~\ref{fig5}~(a), 
which we call a nonmeandering CIC.
At the critical value $\varepsilon\approx 0.011758$,
invariant manifolds of hyperbolic orbits of two chains of the 1:1 resonance 
islands on both sides of the CIC connect, and after that the CIC becomes 
a meandering curve of the first order (Fig.~\ref{fig5}~(b)) and period
$T$. The period of CIC's meandering is simply a period of nearby main
islands \cite{Shinohara97, Simo98}.
At the next critical value $\varepsilon=0.01178721$,
reconnection of invariant manifolds of secondary resonance islands takes place.
The corresponding second-order meandering CIC with period
$79T$ is shown in Fig.~\ref{fig5}~(c).
Highly meandering CICs of higher orders appear with 
further increasing the perturbation amplitude (Fig.~\ref{fig5}~(d)).

Some smooth invariant curves inside the CTB break down under the
perturbation (\ref{eqn3}), and chains  of ballistic resonance islands
appear at their place. Those islands appear in pairs to the north
and south from a CIC due to the flow symmetries
(\ref{*}) and (\ref{**}) (see Fig.~\ref{fig5}~(a) and (b)). Geometry of the CIC, size
and number of the islands, and topology of  their invariant manifolds change with 
variation of the perturbation
amplitude $\varepsilon$ and frequency $\nu$ in a very
complicated way.

\begin{figure}[!htbp]
\centerline{\includegraphics[width=3.4in]{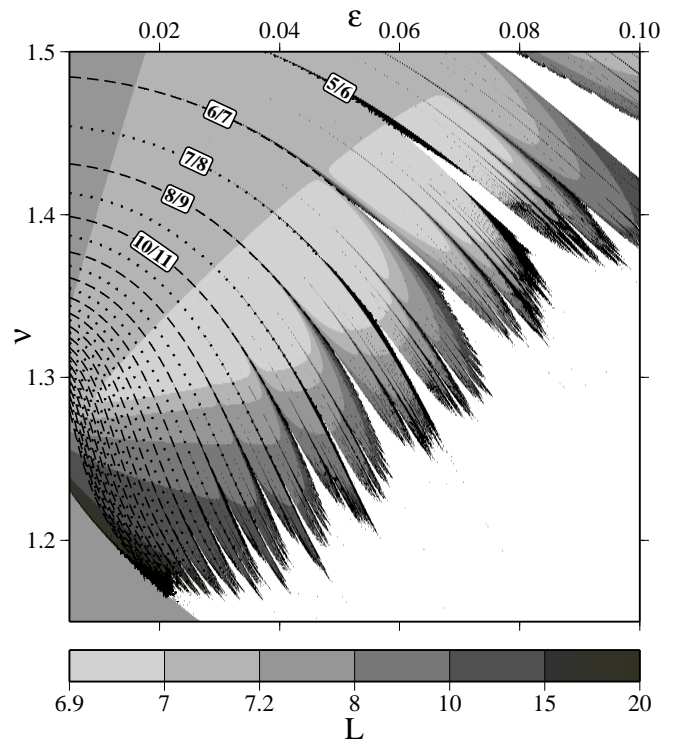}}
\caption{Bird-wing diagram showing the length $L$ of the CIC
by nuances of the grey color in the parameter space $(\varepsilon,\nu)$.
White zone: regime with a broken CTB and cross-jet transport.
Black color: regime 
with a broken CIC but a remaining CTB preventing cross-jet transport.
Resonant bifurcation curves, along which the CIC winding numbers 
$w$ are rational, end up in the dips of ``the wing''. 
Dotted and dashed lines correspond to even and odd resonances, respectively.}
\label{fig6}
\end{figure}

In Fig.~\ref{fig6} we plot in the parameter space the values of the CIC
length $L$ coding it by
nuances of the grey color. White color corresponds to
those values of the parameters $\varepsilon$ and $\nu$
for which cross-jet transport exists due to destruction of the CTB. 
Black color codes the regime with a broken CIC but a remaining CTB
that prevents cross-jet transport ($d_4>1.05$). Dotted and dashed 
lines on the plot are the resonant bifurcation curves along which 
the CIC winding number $w$ is rational. The $m/n$ resonant bifurcation curve 
is the set of values of the control parameters for which a reconnection 
of invariant manifolds of the $n:m$ resonances takes place. The dotted lines
correspond to even resonances with $w=~(2k~-~1)~/~2k$
and the dashed lines are odd resonances with $w=2k/(2k+1)$, $k=1,2,\dots$. 
All those curves end up in the dips of the bird-wing diagram.     
\begin{figure}[!htbp]
\includegraphics[width=3.4in]{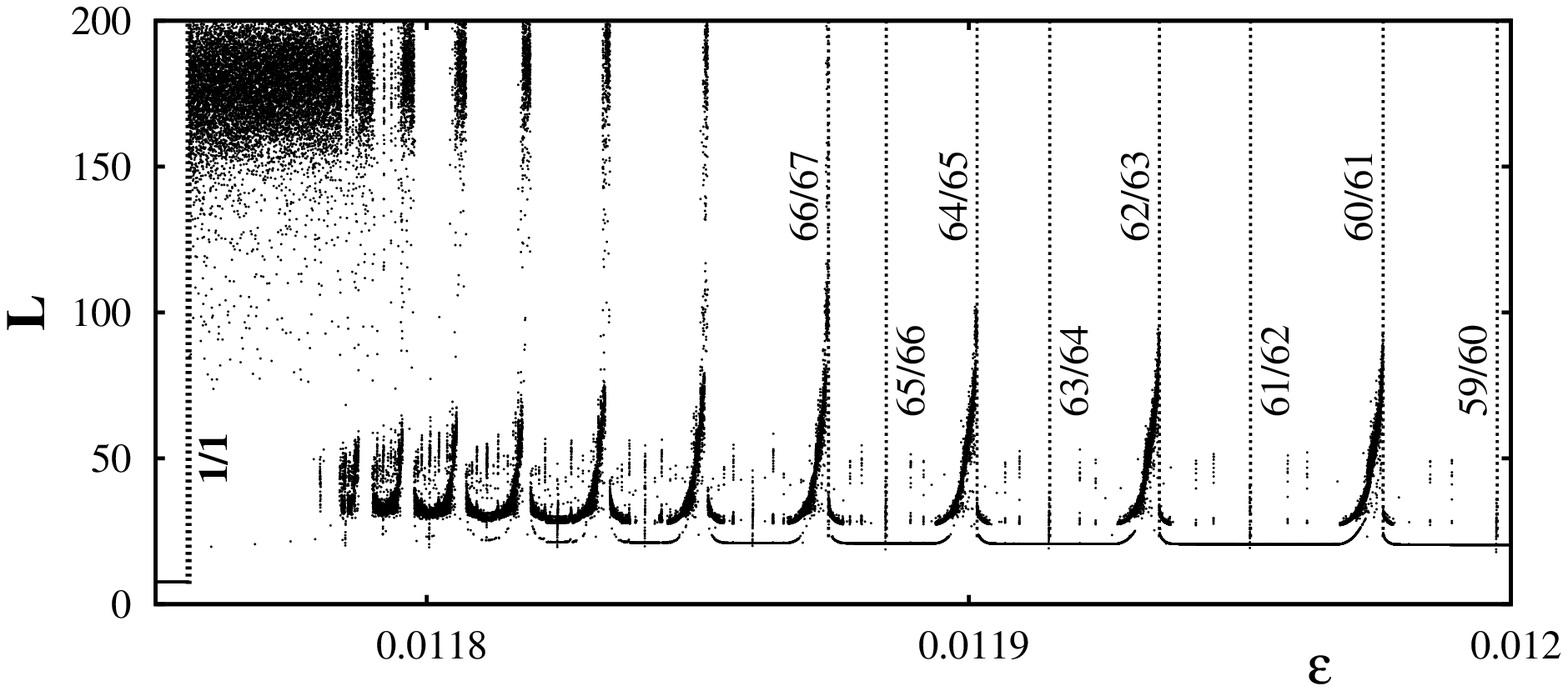}\\
\includegraphics[width=3.4in]{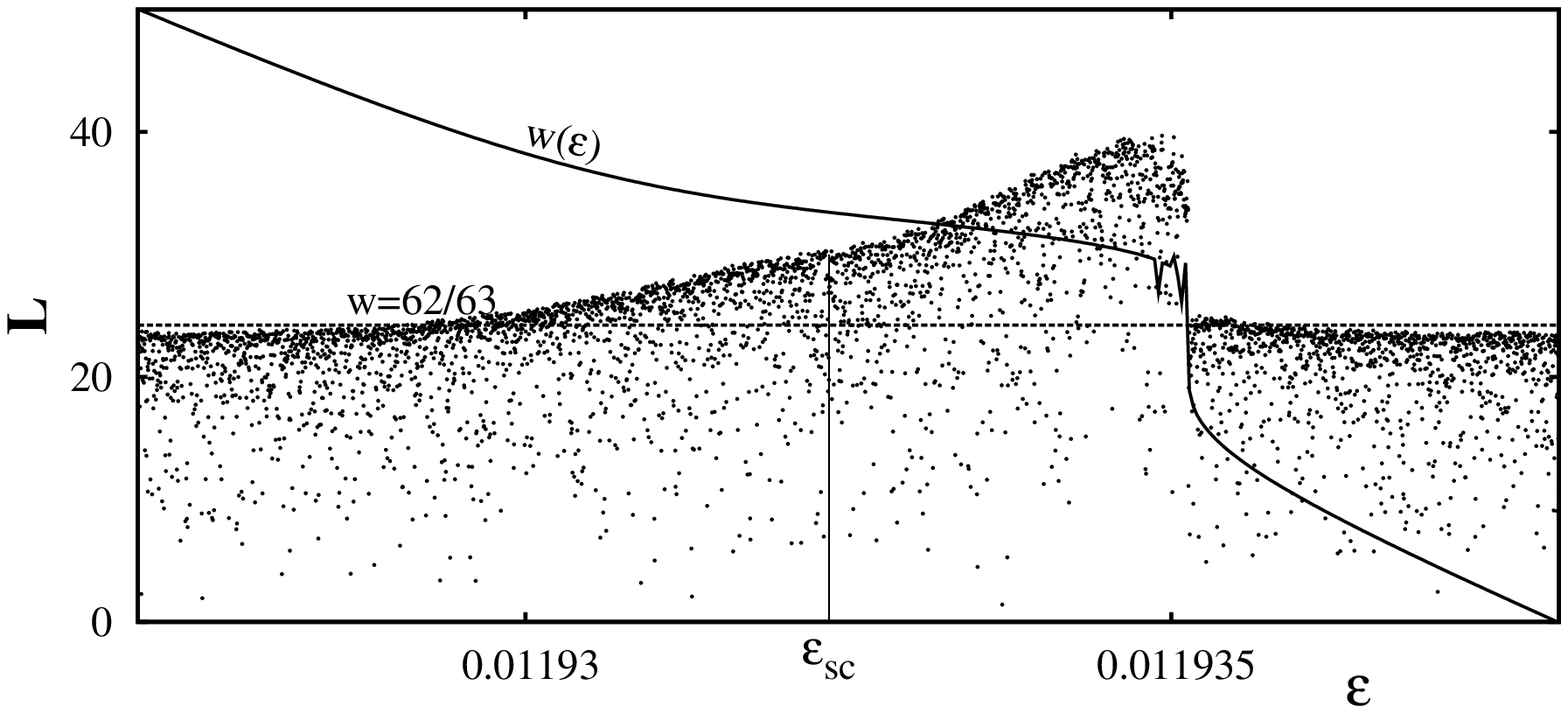}
\caption{Dependence of the length of CIC $L$ on the 
perturbation amplitude at the fixed frequency $\nu=1.2$.
(a) General view of the dependence $L(\varepsilon)$ in the range of
interest $\varepsilon=[0.01175:0.012]$.
Vertical dotted lines correspond to rational values
of the CIC winding number $w$. The resonance $1/1$ appears at 
$\varepsilon=0.11756$. The arrangement of the spikes is explained in the text.
(b) Magnification of one of the wide spikes in panel (a).
Solid line is a winding number profile
$w(\varepsilon)$ with the value $w=62/63$ shown by the dashed line.}
\label{fig7}
\end{figure}

In order to analyze a fractal-like boundary of the bird-wing diagram 
in Fig.~\ref{fig6}, we cross it
horizontally at the frequency $\nu=1.2$ and consider 
the plot $L(\varepsilon)$ in the range of interest of 
$\varepsilon$ (Fig.~\ref{fig7}~(a)). The perturbation frequency $\nu=1.2$
is close to the maximal frequency $f_{\text{max}}$
of particles in the middle of
the jet in the unperturbed flow (see Fig.~\ref{fig1}~(b)). The plot
$L(\varepsilon)$ consists of a number of spikes with different height 
and width. In the range of small 
values of the perturbation amplitude ($\varepsilon<0.011756$),
the length of the CIC is approximately the same $L\approx 7.35$
(a small fragment of the function $L(\varepsilon)$ is shown
in Fig.~\ref{fig7}~(a) just to the left from the vertical line $1/1$).
In that range, the CIC is a nonmeandering curve
(see Fig.~\ref{fig5}~(a)) surrounded by smooth invariant curves and
$1:1$  resonance islands with a heteroclinic topology.
The size of those islands is comparable with
the frame size. The width of the CTB, filled by invariant
curves around the CIC, decreases with increasing the
perturbation amplitude $\varepsilon$. 
\begin{figure}[!htbp]
\centerline{\includegraphics[width=3.4in]{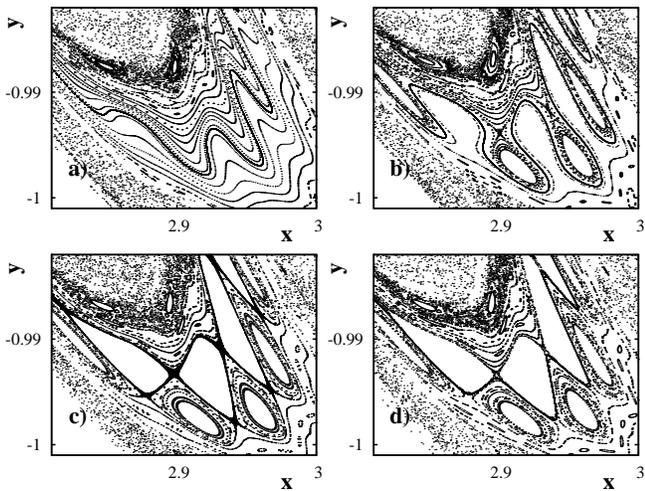}}
\caption{Poincar{\'e} sections  
at $\nu=1.2$ and increasing values of $\varepsilon$ in the
range corresponding to the wide spike with $w=62/63$ in
Fig.~\ref{fig7}~(b). (a) Meandering CIC surrounded by meandering
invariant curves $(\varepsilon=0.011931)$.
(b) CIC meandering between the odd $63:62$ islands  
born as a result of a saddle-center bifurcation ($\varepsilon_{\text{sc}}=0.011934)$.
(c) CIC destruction due to connection of invariant manifolds
of the $63:62$ islands $(\varepsilon=0.01193511)$. A narrow  
stochastic layer appears at the place of the CIC.
(d) CIC appears again $(\varepsilon=0.0119352)$.}
\label{fig8}
\end{figure}

The CIC winding number $w$ changes under a variation of the perturbation
amplitude. At $\varepsilon\simeq 0.011756$, invariant manifolds
of the 1:1 resonance connect, and a central stochastic layer
appears at the place of the CIC. This layer exists up to
$\varepsilon\simeq 0.011785$ (see a random set of points
in Fig.~\ref{fig7}~(a) in that range of $\varepsilon$).
At $\varepsilon> 0.011785$, the CIC appears again. Now
it is a meandering curve of the first order (see Fig.~\ref{fig5}~(b))
whose length is larger due to reconnection of the 1:1
resonance islands. As $\varepsilon$ increases further,
the CIC length $L$ changes in a wide range.
Smooth fragments with approximately the same value of $L\approx 20$
alternate with spikes of different height and width.
The spikes are condensed, when approaching to the value
$w=1/1$, and overlap in the range $\varepsilon\simeq [0.011756:0.011785]$.

\begin{figure}[!htbp]
\centerline{\includegraphics[width=3.4in]{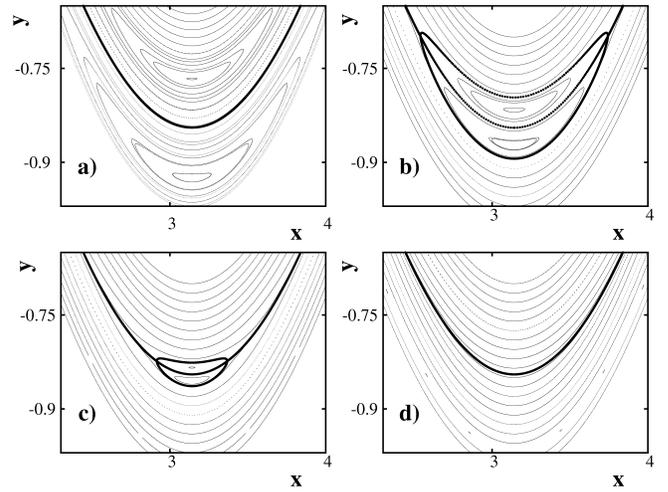}}
\caption{Poincar{\'e} sections at $\varepsilon=0.015$ and increasing
values of $\nu$.
(a) CIC between islands of the even 2:1 resonance
($\nu=2.51<2f_{\text{max}}=2.556$). (b) Reconnection of invariant manifolds of 
that resonance and a formation of a vortex pair
with a narrow stochastic layer shown by bold curves 
($\nu=2.55$). (c) The vortex size decreases with increasing $\nu$
($\nu=2.555$). (d) Past some critical value of $\nu$, the vortex pair
disappears and CIC appears again ($\nu=2.56$).}
\label{fig9}
\end{figure}

The arrangement of the spikes in Fig.~\ref{fig7}~(a) can be explained using a 
representation of rational numbers by continued fractions. A continued fraction
is the expression 
\begin{equation}
c=[a_0;a_1,a_2,a_3,\dots]=a_0+\frac{1}{a_1+\dfrac{1}{a_2+\dfrac{1}
{a_3+\cdots}}},
\end{equation}
where $a_0$~is an integer number and the other $a_n$~are natural 
numbers. Any rational (irrational) number can be represented by 
a continued fraction with a finite (infinite) number of elements.  
The spikes in Fig.~\ref{fig7}~(a) are arranged in convergent series in such a way 
that each spike in a series generates a series of spikes of the next order. 
For example, the series of the integer $n:1$ resonance has the winding numbers 
equal to $1/n$ or $[0;n]$ in the continued-fraction representation. 
Each spike in that series generates a series of resonance spikes of the 
next order converging to the parent spike. Winding numbers of those resonances 
are $[0;n,i]$, $i=2,3,4\dots$ (at $i=1$, one gets a spike in the main series
because of the identity $[a_0;a_1,\dots, a_n,1]\equiv[a_0;a_1,\dots, a_n+1]$). 
The spikes with $[0;1,i]=i/(i+1)$ converge to  
the spike of the $1:1$ resonance. That is clearly seen in Fig.~\ref{fig7}~(a). 
The direction of convergence of the spikes in a series alternate with the series 
order: the winding number increases in the series of the first order, 
decreases in the series of the second order, and increases again in the series of 
the third order. That is why a chaotic region in  Fig.~\ref{fig7}~(a)  is situated 
to the right from the $1:1$ resonance, i.~e. in the range of smaller values of 
$w$. Whereas, it is to the left for  the series of the second order, i.~e. 
in the range of larger values of $w$. There also exists an additional 
hierarchical structure with fractional $1:n$ resonances, whose frequencies are 
below the $1:1$ resonance frequency, and  a series with resonances 
corresponding to 
the spikes below $[0;1,i,(1)]$, for example, a clearly visible series of spikes 
below $[0;1,4,(1)]$ converging to the spike $5/6$ in the $L$ diagram
(see Fig.~\ref{fig6}). 

Unfortunately, we could not identify 
series of the third and a higher order because  numerical errors in 
identifying the winding numbers are greater than 
the distance between the spikes of higher-order series. Unresolved regions
on the plot $L(\varepsilon)$ in Fig.~\ref{fig7}~(a) appear 
because of a decrease of the distance between the spikes in the same series with 
increasing series number, a process resembling Chirikov's overlapping of 
resonances.  

A magnification of one of the wide spikes is shown in Fig.~\ref{fig7}~(b).
We plot the winding number profile $w(\varepsilon)$ together 
with the function $L(\varepsilon)$ for the spike.
To illustrate what happens with the CIC and its surrounding with 
increasing $\varepsilon$ we plot the corresponding Poincar{\'e}
section in Fig.~\ref{fig8}. In the range $\varepsilon\simeq[0.0119:0.01192]$
the lengths of the CIC and surrounding invariant curves increase slowly
due to small changes in their geometry (Fig.~\ref{fig8}~(a)).
After a saddle-center bifurcation at $\varepsilon_{\text{sc}}\simeq0.011934$,
there appear two chains of homoclinically connected 
$63:62$ islands separated by a meandering CIC (Fig.~\ref{fig8}~(b)).
The amplitude of the CIC meanders increases with further increasing 
$\varepsilon$ in the range $\varepsilon\simeq [0.011934:0.011935]$.
In that range the CIC disappears and appears again in a random-like
manner (see the corresponding fragment on the plot $L(\varepsilon)$
in Fig.~\ref{fig7}~(b)) due to overlapping of higher-order
resonances and reconnection of their invariant manifolds.
The example of such a reconnection for the $63:62$ 
resonance at $\varepsilon=0.01193511$ is shown in Fig.~\ref{fig8}~(c)
where a stochastic layer appears at the place of the CIC.
As $\varepsilon$ increases further, the CIC appears again but with a smaller number of
meanders (Fig.~\ref{fig8}~(d)). Animation of the corresponding patterns is 
available at http://dynalab.poi.dvo.ru/papers/cic.avi.

The other wide spikes in the
plot $L(\varepsilon)$ with a similar structure are caused by another 
odd resonances between the external perturbation
and particle's motion along the CIC.
Under a CIC resonance with the winding number $w=m/n$, we mean
reconnection of invariant manifolds of the resonance $n:m$ and
onset of a local stochastic layer.
The narrow spikes, situated between the wide ones in
Fig.~\ref{fig7}~(a), correspond to reconnection 
of even resonances. They are hardly resolved on the plot.
Even resonances of higher orders have a smaller effect on CIC
geometry then odd resonances. As an example, we illustrate in Fig.~\ref{fig9}
metamorphosis of the CIC with the winding number $w=1/2$.
The perturbation amplitude is  fixed at a rather small value
$\varepsilon=0.015$ and the frequency increases in the range
$2.51<\nu<2.556=2f_{\text{max}}$.
At $\nu=2.51$ there are islands of the even $2:1$ 
resonance separated by a CIC (Fig.~\ref{fig9}~(a)).
At some critical value of $\nu$ invariant manifolds of the $2:1$
resonance connect and the islands form a tight vortex-pair
structure surrounded by a narrow stochastic layer (see Fig.~\ref{fig9}~(b)
at $\nu=2.55$). The size of the pair decreases gradually 
with further increasing $\nu$, and the corresponding hyperbolic orbits approach each other 
(see Fig.~\ref{fig9}~(c) at
$\nu=2.555$). At some critical value of $\nu$,
hyperbolic and elliptic orbits of the resonance collide
and annihilate, and CIC appears again (see Fig.~\ref{fig9}~(d)
at $\nu=2.56$). Vortex pairs of the other even resonances are formed in a similar way. 
The higher is the order of the resonance, the smaller is the vortex size.

\begin{figure}[!htbp]
\centerline{\includegraphics[width=3.4in]{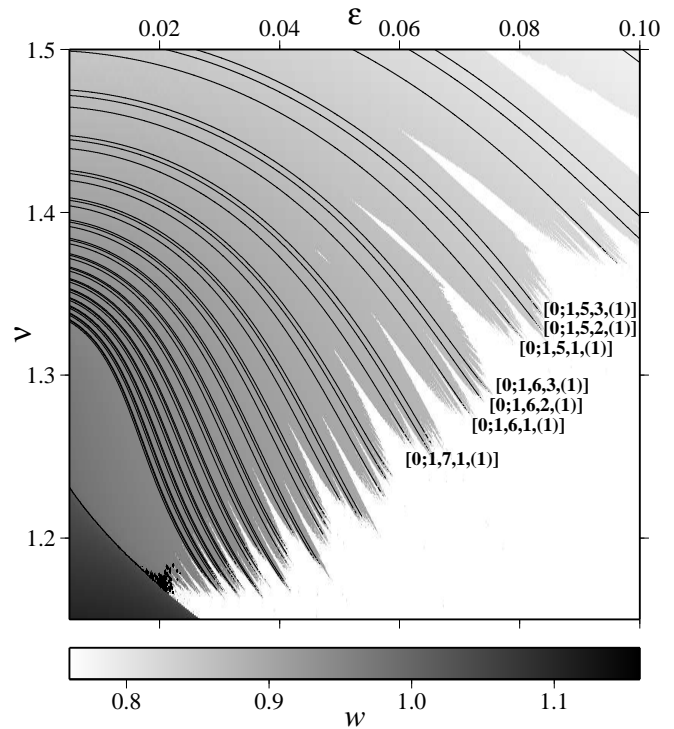}}
\caption{Bird-wing diagram $w(\varepsilon,\nu)$ in the parameter space 
showing values of the winding number $w$ of the CIC by nuances of the grey 
color. White zone: regime with broken CTB and cross-jet transport. The curves 
with irrational winding numbers end up on the tips of the
``feathers of the wing'' (some of them are marked by the corresponding
noble numbers), whereas the curves with rational winding numbers
(shown in Fig.~\ref{fig6}) end up in the dips of ``the wing''.}
\label{fig10}
\end{figure}

We conclude this section by computing winding numbers $w$ of the CIC in the 
parameter space. The result is shown in the bird-wing diagram 
in Fig.~\ref{fig10}. The CIC does not exist in the 
white region where the CTB is broken and cross-jet transport takes place. 
The curves, which end up on the tips 
of the ``feathers of the wing'', have winding numbers $w$ 
with the following continued-fraction representation: 
$[a_0;a_1,\dots, a_n,(1)]$. These are so-called noble numbers which 
are known to be the numbers that cannot be approximated by continued-fraction
sequences to better accuracy than the so-called Diophantine condition
(see, for example, \cite{Almeida}). The CICs 
with noble winding numbers are in a sense the most structurally robust 
invariant curves, i.~e. they may survive under a comparatively large perturbation
preventing cross-jet transport. The noble curves are arranged in series 
like the resonant bifurcation curves with rational 
winding numbers which end up in the dips of ``the wing''
in the bird-wing diagram in Fig.~\ref{fig6}.
For example, the noble series $[0;1,i,(1)]$ 
in Fig.~\ref{fig10} corresponds to the resonance series $[0;1,i]$ 
in Fig.~\ref{fig6}. In the $w$ diagram we show a few representatives 
of the noble series $[0;1,i,(1)]$ and series of the next order 
$[0;1,i,j,(1)]$ (see Fig.~\ref{fig10} with $j=2,3$).

\section{Breakdown of central transport barrier}
\begin{figure}[!htbp]
\centerline{\includegraphics[width=3.4in]{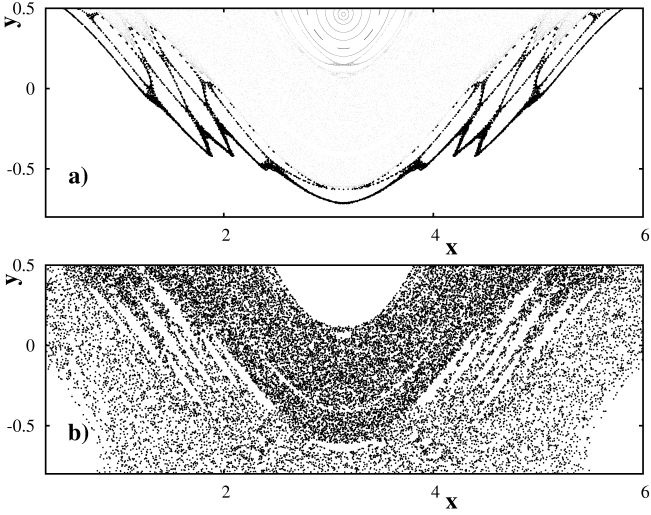}}
\caption{Destruction of central transport barrier upon moving in the parameter 
space along a resonant bifurcation curve with the rational winding number 
$w=8/9$. (a) Narrow stochastic layer on the 
Poincar{\'e} section is confined between invariant curves which provide 
a transport barrier. The perturbation parameters 
($\varepsilon = 0.04889$, $\nu=1.31625$) are chosen on the 
curve with $w=8/9$ nearby its right edge (see Fig.~\ref{fig6}). 
(b) Onset of cross-jet transport at the values of 
parameters ($\varepsilon = 0.054$, $\nu=1.285$) chosen 
in the white zone of that dip.}
\label{fig11}
\end{figure}
\begin{figure}[!htbp]
\centerline{\includegraphics[width=3.4in]{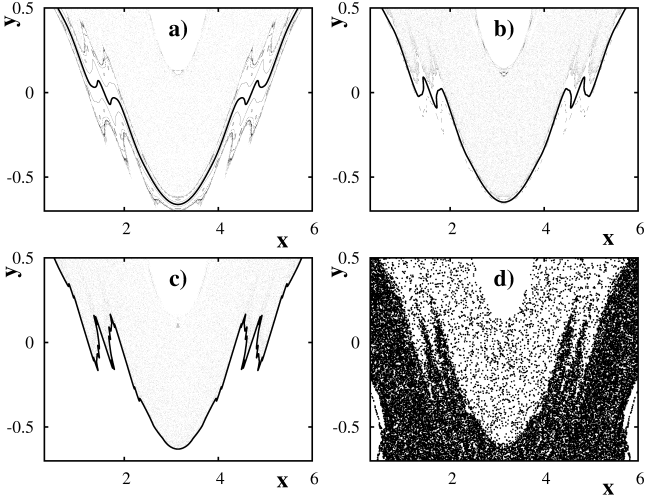}}
\caption{Destruction of central transport barrier upon moving in the 
parameter space along a curve 
with the noble value of the CIC's winding number $w=[0;1,5,1,(1)]$. 
When approaching a tip of the corresponding ``feather of the wing'' 
in Fig.~\ref{fig10}, one observes on the Poincar{\'e} section a   
decrease in the width of transport barrier with  a CIC (bold curves) inside. 
(a) $\varepsilon=0.07017$, $\nu=1.367875$. (b) $\varepsilon=0.07416$, 
$\nu=1.350375$. (c) $\varepsilon=0.0796067$, $\nu=1.325875$. 
(d) Onset of cross-jet transport at the values of 
parameters ($\varepsilon = 0.08$, $\nu=1.3142$) chosen beyond a tip of that 
``feather'' in the white zone in Fig.~\ref{fig10}.}
\label{fig12}
\end{figure}

We have studied in the preceding section properties of the CIC 
which has been shown to be a diagnostic means to characterize CTB and 
its destruction. CTB separates water masses to the south and 
the north from the central jet and prevents their mixing. It is not a homogeneous jet-like 
layer but consists of chains of ballistic islands, narrow stochastic layers, 
and meandering invariant curves of different orders and periods 
(including a CIC) to be confined by invariant curves from the south and 
the north. Those curves break down one after another when  increasing 
the perturbation amplitude $\varepsilon$,  producing stochastic layers at 
their place on both sides of the central
jet, until the stochastic layers merge with one another and with 
stochastic layers around the southern and northern circulation cells producing 
a global stochastic layer and onset of cross-jet transport. 

Upon moving along any resonant bifurcation curve with a rational value 
of  the winding number $w$ in the bird-wing diagram in Fig.~\ref{fig6}, we 
have those values of the perturbation amplitude $\varepsilon$ and 
frequency $\nu$ at which the corresponding CIC is broken due to reconnection 
of invariant manifolds. It does not mean that CTB is broken as well. That is 
the case only if we are at the dips of the ``wing''. 
The process of CTB destruction for this type of movement in the parameter 
space is illustrated in Fig.~\ref{fig11}. We fix 
a point ($\varepsilon = 0.04889$, $\nu=1.31625$) on the resonant bifurcation curve 
with $w=8/9$ nearby its right edge in Fig.~\ref{fig6} and plot the 
corresponding Poincar{\'e} 
section. A narrow stochastic layer, confined between invariant 
curves providing a transport barrier,  appears on the Poincar{\'e} section 
in panel (a) at the place of a broken CIC. The barrier will be broken if 
one would choose the values of parameters in the white zone in Fig.~\ref{fig6}. 
Merging of southern and northern stochastic layers and 
onset of cross-jet transport are shown in panel (b) at $\varepsilon = 
0.054$, $\nu=1.285$.

Upon moving along any curve with a noble value 
of the winding number $w$ in the bird-wing diagram in Fig.~\ref{fig10}, we 
have those values of the perturbation amplitude $\varepsilon$ and 
frequency $\nu$ at which a CIC with the corresponding noble number 
exists. The process of CTB destruction for the motion in the parameter 
space along the noble curve with $w=[0;1,5,1,(1)]$ 
is illustrated in Fig.~\ref{fig12}. When moving 
to the tip of the corresponding ``feather of the wing'' in Fig.~\ref{fig10}, 
one observes progressive destruction of invariant curves and decrease of the 
width of the transport barrier (panels (a) and (b)) unless a single 
CIC remains as the last barrier to cross-jet transport (panel (c)).
Onset of cross-jet transport (panel (d)) happens  at the values 
of parameters chosen beyond a tip of that 
``feather'' in the white zone in Fig.\ref{fig10}.

Upon moving along any resonant bifurcation curve to the corresponding  dip 
of the bird-wing diagram in Fig.~\ref{fig6}, we find cross-jet transport  
at smaller values of the perturbation amplitude as compared to the case with   
irrational winding numbers because in order to provide cross-jet transport 
in the first case it is enough to destruct all the KAM curves. Whereas, CICs with 
irrational and especially noble values of the winding number may deform 
in a complicated way but still survive under increasing $\varepsilon$ up to 
comparatively large values.

\section{Conclusion}

Being motivated by the problem of cross-jet transport in geophysical flows in the 
ocean and atmosphere, we have studied in detail topology of a central transport 
barrier (CTB) and its destruction in a simple kinematic model of a meandering 
current with chaotic advection of passive particles (Fig.~\ref{fig1}) 
that belong to the class of 
non-degeneracy Hamiltonian systems. Direct computation of the
amplitude--frequency 
diagram (Fig.~\ref{fig2}) demonstrated onset of cross-jet transport at surprisingly 
small values of the perturbation amplitude $\varepsilon$ provided that 
the perturbation frequency $\nu$ was sufficiently large. As an indicator of the 
strength of the CTB and its topology, we used a  
central invariant curve (CIC) which was constructed by iterating 
indicator points by a numerical procedure borrowed from theory of nontwist maps. 
The CTB has been shown to exist provided a set of the iterations was bounded. 
Otherwise, cross-jet transport has been observed (Fig.~\ref{fig3}). 
The results were presented as a diagram of the box-counting dimension of those sets 
of iterations in Fig.~\ref{fig4}. 

Geometry of the CIC has been shown to be  highly sensitive to 
small variations in the parameters near a fractal-like boundary of the diagram 
(Fig.~\ref{fig5}). Quantifying complexity of the CIC's form by its length $L$, 
we computed the corresponding $L$ diagram looking like a bird wing with 
a fractal-like boundary (Fig.~\ref{fig6}).
Resonant bifurcation curves with rational 
winding numbers $m/n$ end up in the dips of the boundary. Along those curves 
in the parameter space,  
invariant manifolds of the corresponding $n:m$ resonances connect providing a 
destruction of the CIC. Scenarios of the reconnection are different for odd 
(Fig.~\ref{fig8}) and even (Fig.~\ref{fig9}) resonances. Animation of the 
process at http://dynalab.poi.dvo.ru/papers/cic.avi provides a visual 
demonstration of complexity of topology of the CTB and its destruction. 
Computing the winding number $w$ of the CIC, we have got an information about 
those values of the perturbation parameters at which the CTB is strong or weak. 
Using representation of the values of winding numbers by continued fractions, 
we were able to order spikes with rational values of $w$ in Fig.~\ref{fig7}
into hierarchical series of the corresponding CIC resonances. 
The curve, which end up on the tips of ``feathers of the wing'' in the 
winding-number diagram in Fig.~\ref{fig10}, have noble winding numbers
which are so irrational that the corresponding CICs break down in the 
last turn when varying the perturbation parameters. The noble curves 
have been found to be arranged in series like the resonant bifurcation curves 
with rational values of $w$. Destruction of CTB is illustrated for two ways 
in the parameter space: upon moving along resonant bifurcation curves with 
rational values of $w$  (Fig.~\ref{fig11}) and along curves with  
noble values of $w$ (Fig.~\ref{fig12}). 

In conclusion we address two points that may be important in possible
aplications of the results abtained.
Molecular diffusion in laboratory experiment and turbulent diffusion
in geophysical flows are expected to wash out ideal fractal-like
structures caused by chaotic advection after a characteristic
time scale. The question is what is this scale. As to molecular diffusion
in the ocean, the diffusion time-scale $L^2/D$ is very large
since the diffusion coefficient is of the order of
$D\simeq 10^{-5}$ cm${}^2$/sec and $L$ is of a kilometer scale.
The scale of molecular diffusion in laboratory tanks is, of course,
much smaller. However, some fractal-like structures have been 
observed in real laboratory experiments (see, for example,
Ref.~\cite{SKG96} and the book \cite{Book08} for a recent review
of experiments).
Modelling of a combined effect of chaotic advection and
turbulent diffusion in the ocean is a hard problem
deserving a special consideration. Any kind of diffusion
is expected to intensify cross-jet transport.

The advantage of the kinematic approach is its ability to identify
different factors that may 
enhance or suppress cross-jet transport.
However, the results obtained with our simplified kinematic model
should be taken with caution to describe cross-jet transport
in real geophysical flows. In any kinematic model the velocity
field is postulated based on known features of the current
while in dynamic models it should obey dynamical equations
following from the conservation of potential
vorticity \cite{P87,P91}. It is very difficult to
formulate an analytic and dynamically consistent model
with chaotic advection (for a discussion and examples of 
such models see \cite{DM93,P.H.Haynes,PK06,KSD08}).
One approach is to seek solutions of the
fluid dynamics equations that are self-consistent to linear
order \cite{Lipps}.
Some aspects of cross-jet transport in a 
linearized model with a zonal Bickley jet current and two
Rossby waves have been studied in Refs.~\cite{DM93,PL95,Rypina}.
Potential vorticity is not exactly conserved within the
linear approximation but models that are self-consistent to
linear order provide a compromise between the self-consistency
demands and the fruitfulness of Hamiltonian models.

The question, how predictions of kinematic 
models in destruction of barriers to cross-jet transport carry over to more 
realistic dynamical models, remains open. We plan in the future 
to apply the methods developed 
in the present paper to a dynamically consistent model of a meandering current 
with Rossby waves.  

\section*{Acknowledgments}
The work was supported partially by the Program
``Fundamental Problems of  Nonlinear Dynamics'' of the Russian
Academy of Sciences and by the Russian Foundation  
for Basic Research (project no. 09-05-98520).

\end{document}